\documentclass[11pt,amsmath,amssymb,aps,prd,amssymb,showpacs,nobibnotes]
              {revtex4-2}

\usepackage{amsmath,amssymb,tikz,hyperref,empheq}
\usepackage{graphicx,psfrag}
\usepackage{hyperref} 
\usepackage{dsfont}    
\usepackage{color}

\newcommand\trick[1]{}
\newcommand{\be}{\begin{equation}} 
\newcommand{\ee}{\end{equation}}
\newcommand{\eq}[1]{(\ref{#1})}
\newcommand{\bit}{\begin{itemize}}  \newcommand{\eit}{\end{itemize}}
\newcommand{\ben}{\begin{enumerate}}  \newcommand{\een}{\end{enumerate}}

\newcommand{\bra}[1]{\langle #1|}
\newcommand{\ket}[1]{|#1 \rangle}

\newcommand{\rf}[1]{(\ref{#1})}

\def\bd{\begin{document}}
\def\ed{\end{document}}
\def\bea{\begin{eqnarray}}
\def\eea{\end{eqnarray}}
\let\bm=\bibitem

\def\la{\langle}
\def\ra{\rangle}

\def\npb#1#2#3{Nucl. Phys. {\bf{B#1}} #3 (#2)}
\def\plb#1#2#3{Phys. Lett. {\bf{#1B}} #3 (#2)}
\def\prl#1#2#3{Phys. Rev. Lett. {\bf{#1}} #3 (#2)}
\def\prd#1#2#3{Phys. Rev. {D bf{#1}} #3 (#2)}
\def\cmp#1#2#3{Comm. Math. Phys. {\bf{#1}} #3 (#2)}
\def\cqg#1#2#3{Class. Quantum Grav. {\bf{#1}} #3 (#2)}
\def\nppsa#1#2#3{Nucl. Phys. B (Proc. Suppl.) {\bf{#1A}}#3 (#2)}
\def\ap#1#2#3{Ann. of Phys. {\bf{#1}} #3 (#2)}
\def\ijmp#1#2#3{Int. J. Mod. Phys. {\bf{A#1}} #3 (#2)}
\def\rmp#1#2#3{Rev. Mod. Phys. {\bf{#1}} #3 (#2)}
\def\mpla#1#2#3{Mod. Phys. Lett. {\bf A#1} #3 (#2)}
\def\jhep#1#2#3{J. High Energy Phys. {\bf #1} #3 (#2)}
\def\atmp#1#2#3{Adv. Theor. Math. Phys. {\bf #1} #3 (#2)}

\def\sst{\scriptscriptstyle}
\def\thetabar{\bar\theta}
\def\Tr{{\rm Tr}}
\def\one{\mbox{1 \kern-.59em {\rm l}}}

\def\a{\alpha}      \def\da{{\dot\alpha}}  \def\dA{{\dot A}}
\def\b{\beta}       \def\db{{\dot\beta}}
\def\g{\gamma}  \def\G{\Gamma}  \def\dc{{\dot\gamma}}
\def\d{\delta}  \def\D{\Delta}  \def\ddt{\dot\delta}
\def\e{\epsilon}
\def\ve{\varepsilon}
\def\uve{\upvarepsilon}
\def\f{\phi}    \def\F{\Phi}    \def\vvf{\f}
\def\vphi{\varphi}
\def\h{\eta}
\def\k{\kappa}
\def\l{\lambda} \def\L{\Lambda}
\def\m{\mu} \def\n{\nu}
\def\o{\omega}
\def\p{\pi} \def\P{\Pi}
\def\r{\rho}
\def\s{\sigma}  \def\S{\Sigma}
\def\t{\tau}
\def\th{\theta} \def\Th{\Theta} \def\vth{\vartheta}
\def\X{\Xeta}
\def\z{\zeta}

\def\na{\nabla}

\def\cA{{\cal A}} \def\cB{{\cal B}} \def\cC{{\cal C}}
\def\cD{{\cal D}} \def\cE{{\cal E}} \def\cF{{\cal F}}
\def\cG{{\cal G}} \def\cH{{\cal H}} \def\cI{{\cal I}}
\def\cJ{{\mathscr J}} \def\cK{{\cal K}} \def\cL{{\cal L}}
\def\cM{{\cal M}} \def\cN{{\cal N}} \def\cO{{\cal O}}
\def\cP{{\cal P}} \def\cQ{{\cal Q}} \def\cR{{\cal R}}
\def\cS{{\cal S}} \def\cT{{\cal T}} \def\cU{{\cal U}}
\def\cV{{\cal V}} \def\cW{{\cal W}} \def\cX{{\cal X}}
\def\cY{{\cal Y}} \def\cZ{{\cal Z}}
\def\ct{{\cal t}}

\def\ua{\underline{\alpha}}
\def\uc{\underline{\phantom{\alpha}}\!\!\!\gamma}
\def\um{\underline{\mu}}
\def\ud{\underline\delta}
\def\ue{\underline\epsilon}
\def\una{\underline a}\def\unA{\underline A}
\def\unb{\underline b}\def\unB{\underline B}
\def\unc{\underline c}\def\unC{\underline C}
\def\und{\underline d}\def\unD{\underline D}
\def\une{\underline e}\def\unE{\underline E}
\def\unf{\underline{\phantom{e}}\!\!\!\! f}\def\unF{\underline F}
\def\unm{\underline m}\def\unM{{\underline M}}
\def\unn{\underline n}\def\unN{{\underline N}}
\def\unp{\underline{\phantom{a}}\!\!\! p}\def\unP{\underline P}
\def\unq{\underline{\phantom{a}}\!\!\! q}
\def\unQ{\underline{\phantom{A}}\!\!\!\! Q}
\def\unH{\underline{H}}

\def\As {{A \hspace{-6.4pt} \slash}\;}
\def\bs {{b \hspace{-6.4pt} \slash}\;}
\def\Ds {{D \hspace{-6.4pt} \slash}\;}
\def\Gts {{\Gt \hspace{-6.4pt} \slash}\;}
\def\ds {{\del \hspace{-6.4pt} \slash}\;}
\def\ss {{\s \hspace{-6.4pt} \slash}\;}
\def\ks {{ k \hspace{-6.4pt} \slash}\;}
\def\ps {{p \hspace{-6.4pt} \slash}\;}
\def\xs {{x \hspace{-6.4pt} \slash}\;}
\def\pas {{{p_1} \hspace{-6.4pt} \slash}\;}
\def\pbs {{{p_2} \hspace{-6.4pt} \slash}\;}
\def\cFs {{{\cal F} \hspace{-6.4pt} \slash}\;}
\def\Dss {{D \hspace{-7.5pt} \slash}\;}
\def\dss {{\del \hspace{-7.0pt} \slash}\;}

\def\Ah{{\hat{A}}}
\def\Dh{{\hat{D}}}
\def\Gh{{\hat{G}}}
\def\Fh{{\hat{F}}}
\def\Ih{{\hat{I}}}
\def\Jh{{\hat{J}}}
\def\Kh{{\hat{K}}}
\def\Lh{{\hat{L}}}
\def\Ph{{\hat{P}}}
\def\Rh{{\hat{R}}}
\def\Vh{{\hat{V}}}
\def\Xh{{\hat{X}}}

\def\ah{{\hat{\a}}}
\def\bh{{\hat{\b}}}
\def\gh{{\hat{\g}}}
\def\dh{{\hat{\d}}}
\def\rh{{\hat{\r}}}
\def\hh{\hat{h}}
\def\uh{\hat{u}}
\def\xh{\hat{x}}
\def\yh{\hat{y}}
\def\ph{\hat{p}}
\def\xih{\hat{\xi}}
\def\chih{\hat{\chi}}
\def\Psih{\hat{\Psi}}
\def\phih{\hat{\phi}}

\def\psit{\tilde{\psi}}
\def\Psit{\tilde{\Psi}}
\def\Psibt{\tilde{\bar{Psi}}}

\def\lambdat{\tilde {\lambda}}
\def\st{\tilde{\sigma}}

\def\delt{\tilde{\delta}}
\def\Phit{\tilde{\Phi}}
\def\Phitb{\overline{\tilde{Phi}}}
\def\tht{\tilde{\th}}
\def\lt{\tilde{\l}}
\def\chit{\tilde{\chi}}
\def\phit{\tilde{\phi}}

\def\At{\tilde{A}}
\def\Bt{\tilde{B}}
\def\Ct{\tilde{C}}
\def\Dt{\tilde{D}}
\def\Et{\tilde{E}}
\def\Ft{\tilde{F}}
\def\Gt{\tilde{G}}
\def\Ht{\tilde{H}}
\def\It{\tilde{I}}
\def\Jt{\tilde{J}}
\def\Qt{\tilde{Q}}
\def\Rt{\tilde{R}}
\def\Mt{\tilde{M }}
\def\Nt{\tilde{N}}
\def\St{\tilde{S}}
\def\Tt{\tilde{T}}
\def\Vt{\tilde{V}}
\def\Xt{\tilde{X}}
\def\at{\tilde{a}}
\def\dt{\tilde{d}}
\def\htt{\tilde{h}}
\def\ft{\tilde{f}}
\def\gt{\tilde{g}}
\def\ot{\tilde{\omega}}
\def\pt{\tilde{p}}
\def\qt{\tilde{q}}
\def\vt{\tilde{v}}
\def\nt{\tilde{n}}
\def\ut{\tilde{u}}
\def\wt{\tilde{w}}
\def\zt{\tilde{z}}
\def\xt{\tilde{x}}
\def\yt{\tilde{y}}
\def\Psit{\tilde{\Psi}}
\def\vphit{\tilde{\varphi}}
\def\tD{\tilde{\D}}

\def\eb{\bar{\epsilon}}
\def\delb{\bar{\partial}}
\def\thb{\bar{\theta}}
\def\mub{\bar{\mu}}
\def\lamb{\bar{\l}}
\def\psib{\bar{\psi}}
\def\sb{\bar{\sigma}}
\def\xib{\bar{\xi}}
\def\chib{\bar{\chi}}

\def\Psib{\bar{\Psi}}
\def\Phib{\bar{\Phi}}
\def\Lamb{\bar{\Lambda}}
\def\Sb{{\overline \Sigma}}
\def\cb{\bar{c}}
\def\hb{\bar{h}}
\def\qb{\bar{q}}
\def\wb{\bar{w}}
\def\ub{\bar{u}}
\def\zb{{\bar{z}}}
\def\Hb{\bar{H}}
\def\Qb{{\bar Q}}
\def\Omegab{\overline{\Omega}}
\def\ob{\overline{\omega}}

\def\Ab{{\overline A}} \def\Bb{{\overline B}} \def\Cb{{\overline C}}
\def\Db{{\overline D}} \def\Eb{{\overline E}} \def\Fb{{\overline F}}
\def\Gb{{\overline G}}
\def\Ib{{\overline I}}
\def\Jb{{\overline J}} \def\Kb{{\overline K}} \def\Lb{{\overline L}}
\def\Mb{{\overline M}} \def\Nb{{\overline N}} \def\Ob{{\overline O}}
\def\Pb{{\overline P}}  \def\Rb{{\overline R}}
 \def\Tb{{\overline T}} \def\Ub{{\overline U}}
\def\Vb{{\overline V}} \def\Wb{{\overline W}} \def\Xb{{\overline X}}
\def\Yb{{\overline Y}} \def\Zb{{\overline Z}}

\def\fb{{\overline f}}
\def\gb{{\overline g}}
\def\nb{{\overline n}}
\def\mb{{\overline m}}
\def\lb{{\overline l}}
\def\yb{{\overline y}}

\def\ldel{{\overleftarrow{\del}}}
\def\rdel{{\overrightarrow{\del}}}
\def\ldeldel{{\overleftarrow{\del^2}}}
\def\rdeldel{{\overrightarrow{\del^2}}}
\def\ldelb{{\overleftarrow{\bar{\del}}}}
\def\rdelb{{\overrightarrow{\bar{\del}}}}

\def\ba{{\bf a}}
\def\bk{{\bf k}}
\def\bl{{\bf l}}
\def\bp{{\bf p}}
\def\bq{{\bf q}}
\def\br{{\bf r}}
\def\bt{{\bf t}}
\def\bu{{\bf u}}
\def\bv{{\bf v}}
\def\bx{{\bf x}}
\def\by{{\bf y}}
\def\bA{{\bf A}}
\def\bR{{\bf R}}
\def\bV{{\bf V}}

\def\bz{{\boldsymbol{\zeta}}}

\def\bone{{\bf 1}}

\def\va{{\vec a}}
\def\vk{{\vec k}}
\def\vp{{\vec p}}
\def\vq{{\vec q}}
\def\vx{{\vec x}}
\def\vy{{\vec y}}
\def\vu{{\vec u}}
\def\vv{{\vec v}}
\def \vH{{\vec H}}
\def \vg{{\vec g}}

\def\vs{{\vec \sigma}}
\def\vtau{{\vec \tau}}

\newcommand{\ov}[1]{\overrightarrow{#1}}

\def\frA{\mathfrak{A}}
\def\frB{\mathfrak{B}}
\def\frC{\mathfrak{C}}
\def\frD{\mathfrak{D}}
\def\frE{\mathfrak{E}}
\def\frF{\mathfrak{F}}
\def\frG{\mathfrak{G}}
\def\frH{\mathfrak{H}}
\def\frM{\mathfrak{M}}
\def\frN{\mathfrak{N}}
\def\frR{\mathfrak{R}}
\def\frW{\mathfrak{W}}

\def\fra{\mathfrak{a}}
\def\frb{\mathfrak{b}}
\def\frf{\mathfrak{f}}
\def\frg{\mathfrak{g}}
\def\frh{\mathfrak{h}}
\def\frl{\mathfrak{l}}
\def\frs{\mathfrak{s}}
\def\fri{\mathfrak{i}}
\def\frj{\mathfrak{j}}

\def\ma{\mathfrak{a}}
\def\mg{\mathfrak{g}}
\def\mh{\mathfrak{h}}
\def\mR{\mathfrak{R}}
\def\mN{\mathfrak{N}}

\newcommand{\nn}{{\nonumber}}

\def\d{\delta}\def\D{\Delta}\def\ddt{\dot\delta}

\def\pa{\partial} \def\del{\partial}
\def\xx{\times}
\def\uno{\mbox{1 \kern-.59em {\rm l}}}

\def\trp{^{\top}}
\def\inv{^{-1}}
\def\dag{\dagger}
\def\pr{^{\prime}}

\def\rar{\rightarrow}
\def\lar{\leftarrow}
\def\lrar{\leftrightarrow}

\newcommand{\0}{\,\!}      
\def\one{1\!\!1\,\,}
\def\im{\imath}
\def\jm{\jmath}

\newcommand{\tr}{\mbox{tr}}
\newcommand{\slsh}[1]{/ \!\!\!\! #1}

\newcommand{\1}{\mbox{1}\hspace{-0.25em}\mbox{l}}

\def\vac{|0\rangle}
\def\lvac{\langle 0|}

\def\hlf{\frac{1}{2}}
\def\ove#1{\frac{1}{#1}}
\newcommand{\hot}[1]{\frac{#1}{2}}

\def\Box{\square}
\def\CC {\mathbb{C}}
\def\FF {\mathbb{F}}
\def\RR{\mathbb{R}}
\def\NN{\mathbb{N}}
\def\ZZ{\mathbb{Z}}
\def\bb#1{{\bf #1}}
\def\bcomment#1{}
\def\bfhat#1{{\bf \hat{#1}}}
\def\VEV#1{\left\langle #1\right\rangle}

\newcommand{\ex}[1]{{\rm e}^{#1}} \def\ii{{\rm i}}

\newcommand{\lrbrk}[1]{\left(#1\right)}
\newcommand{\lrsbrk}[1]{\left[#1\right]}
\newcommand{\sfrac}[2]{{\textstyle\frac{#1}{#2}}}

\def\stw{{\sqrt{2}}}

\def\rf {{\rm f}}
\def\ri {{\rm i}}
\def\rj {{\rm j}}
\def\rn {{\rm n}}
\def\rk {{\rm k}}
\def\rl {{\rm l}}
\def\rr {{\rm r}}
\def\rs {{\scriptscriptstyle \rm S}}
\def\rt {{\scriptscriptstyle \rm T}}

\def\rQ {{\scriptscriptstyle \rm \cQ}}
\def\rR {{\scriptscriptstyle \rm \cR}}

\def\cQb{{\cal \Qb}}
\def\cRb{{\cal \Rb}}
\def\cWb{{\cal \Wb}}

\def\fd {{\rm N}}
\def\afd {{\overline{\rm N}}}

\def \II {I\hspace{-.1em}I\hspace{.1em}}
\def \IIA {\mbox{\II A\hspace{.2em}}}
\def \IIB {\mbox{\II B\hspace{.2em}}}
\def \gs {g^s}
\def \ls {\lambda^s}

\def \I {{\cal I}}
\def \qs {q\hspace{-.53em}/\hspace{.15em}}
\def \ks {k\hspace{-.53em}/\hspace{.15em}}
\def \YM {{\mbox{\tiny YM}}}
\def \gym {g_{\YM}}

\def \Lc {\L_c}
\def\IR{\relax{\rm I\kern-.18em R}}
\def \id {{\bf 1}}

\def\cci{\ell}
\def\ccj{\ell'}

\def\bbq{\pmb{q}}
\def\bom{\pmb{\o}}
\def\bJ{\pmb{J}}
\def\bM{\pmb{M}}
\def\bB{\pmb{B}}
\def\bn{\pmb{n}}
\def\bE{\pmb{E}}

\newcommand{\rrr}[1]{\vskip 0.2cm \noindent{\bf #1} ---}

\long\def\symbolfootnote[#1]#2{\begingroup%
\def\thefootnote{\fnsymbol{footnote}}\footnote[#1]{#2}\endgroup}
\long\def\RemarkBox#1{\begin{flushleft}\fbox{\begin{minipage}
{17.5cm}{\bf Remark:} ~#1\end{minipage}}\end{flushleft}}

\newcommand{\nthu}{{\it Department of Physics, National Tsing-Hua
  University,
  Hsinchu 30013, Taiwan}}

\newcommand{\ctc}{{\it
Center of Theory and Computation, 
National Tsing-Hua University, Hsinchu 30013, Taiwan}}

\newcommand{\ncts}{{\it
National Center for Theoretical Sciences, Taipei 10617, Taiwan}}

\begin{document}
\title{Quantum Kerr Black Hole from
  Matrix Theory of Quantum Gravity}

\author{Chong-Sun Chu}
\affiliation{Department of Physics, National Tsing-Hua
  University, Hsinchu 30013, Taiwan}
\affiliation{Center for Theory and Computation, National Tsing-Hua
  University, Hsinchu 30013, Taiwan}
\affiliation{National Center for Theoretical Sciences, Taipei 10617, Taiwan}

\begin{abstract}

  Recently, a quantum mechanical theory of quantum spaces described by
  a large $N$ non-commutative coordinates is proposed as a model for
  quantum gravity\cite{Chu:2024qil}. In this paper, we construct Kerr
  black hole as a rotating noncommutative geometry solution of this
  theory.  Due to rotation, the fuzzy sphere is deformed into a fuzzy
  ellipsoid, which matches exactly the outer horizon of the Kerr black
  hole in the Boyer-Lindquist coordinates.  Together with a
  half-filled Fermi sea, the fuzzy solution reproduces the
  Bekenstein-Hawking entropy as well as the mass and angular momentum
  of the  Kerr black
  hole.  These results provide  support that the proposed quantum mechanics
  of  quantum spaces as a model of quantum
  gravity.
    
\end{abstract}

\maketitle

\section{Introduction}

General relativity is a classical dynamical theory of continuum spacetime.
Classical spacetime is described by coordinate functions $x^\mu(P)$
defined over the spacetime manifold $\cM$ and $P\in \cM$ is any point
(event) in the spacetime.  Einstein theory of spacetime consists of 
the equipment of a metric on $\cM$.  One may allow the use of any
coordinate systems and impose diffeomorphism invariance as a symmetry.
The simplest Lagrangian which is consistent with the diffeomorphism
symmetry is the Hilbert-Einstein action. One then show that this
dynamical theory of spacetime contains gravity by finding solutions of the
theory and by examining their effects on motion of
probes described by geodesic motion.

A particular interesting and puzzling solution is the black
hole. First it is not a consistent solution as it contains singularity
\cite{penrose:1964wq}, meaning that the description of black holes in terms of
a classical spacetime is incomplete. Besides, it exhibits a
Bekenstein-Hawking entropy \cite{Bekenstein:1972tm,Bekenstein:1973ur}
whose microscopic origin is unknown. (see however \cite{Strominger:1996sh}
for success in microstates counting for supersymmetric black holes
in string theory).
Nevertheless, the area dependent
nature of the entropy suggests that quantum gravity is holographic
\cite{tHooft:1993dmi,Susskind:1994vu}. Moreover, naive consideration of black
hole evolution suggests a loss of quantum information
\cite{Hawking:1976ra,Polchinski:2016hrw,Almheiri:2020cfm}.  It is
believed that only in a theory of quantum gravity can these problems
be properly addressed and resolved.

In a recent paper \cite{Chu:2024qil}, a formulation of quantum gravity
in terms of a quantum mechanics of quantum space was proposed. The
proposal was motivated by a number of considerations.  Recent studies
of black hole in the AdS/CFT correspondence \cite{Maldacena:1997re}
has cumulated at the success in obtaining the Page curve behavior
\cite{Page:1993wv,Page:2013dx} of
the entanglement entropy of the Hawking radiation. However, the origin
and nature of island, a new saddle point in the holographic entropy formula 
in the island proposal
\cite{Engelhardt:2014gca,Penington:2019npb,Almheiri:2019psf,Almheiri:2019hni},
is not clear from a fundamental point of view.
Motivated by the curious feature that island
is located just underneath the black hole horizon,
we postulated \cite{Chu:2022ieq} that the quantum states of
black hole responsible for the Bekenstein-Hawking entropy are given by
a thin shell of Bell degrees of fermionic freedom located at the region just
underneath the horizon. This picture was further developed in
\cite{Chu:2023mqi,Chu:2023agv}
into a bottom-up model of quantum black hole where the energy and entropy of
black holes were successfully accounted for.

To make progress, one wants to have a dynamical theory of quantum space 
where a quantum black hole is obtained as a solution. Our phenomenological
model \cite{Chu:2023mqi,Chu:2023agv} suggests the desired theory must contains
fermionic degrees of freedom. Fermions are featured in 
various popular approaches to quantum  gravity such as string/M theory
\cite{Becker:2006dvp},
AdS/CFT \cite{Maldacena:1997re}, BFSS quantum mechanics
\cite{Banks:1996vh,Taylor:2001vb} or IKKT matrix model
\cite{Ishibashi:1996xs}.  Since supersymmetry is always built-in
in these setups, the
fermion content as well as the supersymmetric action are highly
constrained.  We argued in \cite{Chu:2024qil} that there is actually
no essential reason to insist on having supersymmetry if one is
interested in quantum mechanics.  Gathering the ideas discussed above,
we proposed in \cite{Chu:2024qil} a $SU(N)$ quantum mechanics of
noncommutative geometry as a theory of quantum gravity.  The theory
has a Higgs-like bosonic potential which allows for the construction
of nontrivial ``Higgs-vacuum''.
This is possible since we do not assume
supersymmetry. In general, solutions of the theory are given by dynamical
noncommutative geometry.  In \cite{Chu:2024qil}, an interesting
noncommutative geometric solution of fuzzy sphere was found.  Together
with a half-filled Fermi sea residing over the fuzzy sphere, the
solution possesses energy of a Schwarzschild black hole and a
microstates entropy which agrees precisely with the Bekenstein-Hawking
entropy expected for a quantum black hole. It was also shown that the
interaction energy between two fuzzy spheres has the correct
dependence on the Newton constants and masses as required by
gravity. It was conjectured that Newton gravity between static sources
would be reproduced in the large distance limit.

In this paper, we construct a new time dependent solution of the
quantum mechanics. The solution describes a rotating fuzzy
ellipsoid. With a half-filled Fermi sea, the solution has a microstate
counting that reproduces precisely the Bekenstein-Hawking entropy of a
quantum Kerr black hole. Moreover, the
angular momentum of the solution matches precisely
that of the Kerr black hole and
it has an energy matches that of the classical
mass of Kerr black hole, with the leading correction at the order of
$a^4/r_+^4$.  We propose that the fuzzy ellipsoid with a half-filled
Fermi sea as a quantum description of the Kerr black hole in quantum
gravity.  Since a rotating black hole is a much more complicated
system, the fact that our proposed dynamical theory of quantum space
contains also the Kerr black hole provides further support that the
proposed quantum mechanics theory as a model of quantum gravity.

\section{Kerr Metric in General Relativity}

In general relativity,
a black hole with mass $M$ and angular momentum $J$ is described by
the Kerr metric. The Kerr metric can be written in a number of different ways.
See for example \cite{Visser:2007fj} for a brief introduction to the
Kerr spacetime.
In the  Boyer-Lindquist coordinates, the Kerr metric reads
\bea \label{kerr}
      ds^{2} = - \left(1- \frac{2M r}{\rho ^{2}}\right) dt^{2}
      - \frac{4Mar\sin ^{2}\theta}{\rho ^{2}}dtd\phi
      + \frac{\Sigma}{\rho ^{2}}\sin ^{2}\theta d\phi^{2}  
      + \frac{\rho ^{2}}{\Delta}dr^{2} + \rho ^{2}d\theta^{2},
      \eea
where
\bea
       \rho^{2} := r^{2}+a^{2}\cos ^{2}\theta, \quad
        \Delta := r^{2}- 2M r+a^{2}, \quad
 \Sigma:= (r^2+a^2)^2-a^2 \Delta \sin^2 \th.
\eea
The Boyer-Lindquist coordinates is a generalization of the Schwarzschild
coordinates which allows the use of spherical coordinates. In the asymptotic
region $r \to \infty$, we have
\bea
 ds^2 = && 
 - \left[1- \frac{2M}{r} +O\left(\frac{1}{r^3}\right)\right] dt^{2} 
 - \left[\frac{4Ma\sin^2 \th}{r} + O\left(\frac{1}{r^3}\right)\right]
d\phi dt \\
&& + \left[1- \frac{2M}{r} +O\left(\frac{1}{r^2}\right)\right]
\left[dr^2+ r^2 (d\th^2 + \sin^2 \th d\phi^2)\right]. \nn
\eea
This confirms that $M$ is the mass and $J=aM$ is the angular momentum of the
black hole spacetime. It is interesting to note that for $M=0$, the metric
\eq{kerr} is flat and reads
\bea
ds^2  =&& -dt^2 + \frac{r^2+a^2 \cos^2 \th}{r^2+a^2} dr^2
+ (r^2 +a^2 \cos^2\th) d\th^2  +(r^2+a^2)\sin^2 \th d\phi^2. 
\eea
  This is in fact the standard Minkowski space written using
the ``oblate spherical'' coordinates, which is related to the Cartesian
coordinates with the coordinate transformation
\bea \label{cc}
&&x = \sqrt{r^2+a^2} \sin \th \cos \phi,\quad 
y = \sqrt{r^2+a^2} \sin \th \sin \phi, \quad z = r \cos \th.
\eea
It is clear that, for $a\neq 0$, the oblate spherical coordinates
$(r, \th, \phi)$ is different from the standard polar coordinates
for 3-dimensional Euclidean space.

The Kerr metric is a stationary, axially symmetric and asymptotically flat
vacuum solution to the Einstein equation.
It has the
Killing vectors $t^\a = \del x^\a /\del t$ and $\phi^\a = \del
x^\a/\del \phi$.  The metric is invariant under simultaneous inversion
of $t \to -t$ and $\phi \to - \phi$, which is consistent with the fact
that time reversal of a rotating object reverse the rotation of the
object. The metric reveals a curvature singularity
$R^{\a\b\g\d}R_{\a\b\g\d} \sim 1/ \rho^{12}$ at $\rho =0$.  In
addition the metric has a coordinate
singularity at $\D =0$.
For $a \leq M$, the roots of $\D=0$ are real. The larger solution
\be
r_+ = M + \sqrt{M^2 -a^2}
\ee
denotes the position of the outer horizon. That the surfaces $r=r_\pm$ are 
event horizon can be seen from the fact that they are null there:
$g^{\a\b} \del_\a r \del_\b r = g^{rr} = \D/\r^2 =0$.
Examination of the null geodesics near the surface shows that they are
indeed horizons.
Viewed
in the Cartesian coordinates \eq{cc},
the horizon is an elliptical surface
\be \label{kerr-shape}
\frac{x^2+y^2}{r_+^2+a^2} + \frac{z^2}{r_+^2} =1. 
\ee
We also remark that the outer horizon is a Killing horizon since
the time like Killing vector
$\xi^\a := t^\a + \o_H \phi^\a$ becomes null at $r=r_+$, where 
\be
\o_H = \frac{a}{r_+^2+a^2}
\ee
is the angular velocity of the black hole.

For the Kerr metric, the  horizon area is given by
\be
A = 4\pi (r_+^2 + a^2),
\ee
which give rises to the Bekenstein-Hawking entropy
\be \label{S-BH}
S= \frac{A}{4 G} = \frac{ \pi (r_+^2 + a^2)}{G}.
\ee
The mass of the Kerr black hole, written in terms of $r_+$,  is
\be \label{M-GR}
M = \frac{r_+^2 +a^2}{2G r_+}
\ee
and the angular momentum $J$
\be\label{J-GR}
J = a \frac{r_+^2 +a^2}{2G r_+}.
\ee
We will show below that these properties \eq{S-BH},\eq{M-GR} and \eq{J-GR}
of the Kerr black hole are reproduced by a rotating solution in 
our proposed quantum mechanics of quantum space.

\begin{widetext}
\section{Kerr Black Hole from  Quantum Mechanics of Quantum Space}

Consider the proposal of \cite{Chu:2024qil} for a $SU(N)$
quantum mechanical theory of
3-dimensional quantized space defined by the Lagrangian
\be \label{L}
L =  \tr \left[
   \frac{1}{2M_0} \dot{X}^{a2}
   + \frac{M_P}{N^2} \left([X^a,X^b]^2 + 4  X^{a2}\right)
   + i \dot{\psi}^\dag \psi
   - a_2 \frac{M_P}{N^2}\psi^\dag \s^a X^a  \psi \right]
- a_3 r_X M_P ,
\ee
or the Hamiltonian
\be \label{H}
H = \tr\left[
\frac{ M_0}{2} P^{a2}
  - \frac{M_P}{N^2}\left( [X^a,X^b]^2 +4 X^{a2}\right)
+a_2\frac{M_P}{N^2} \psi^\dag \s^a X^a  \psi 
\right]+ a_3 r_X M_P.
\ee
\end{widetext}
Here $X^a_{mn}$, $a=1,2, 3$ represent the coordinates of a
  3-dimensional quantized space,
  $\s^a$ are the Pauli matrices and
  $\psi_{mn}, \psi_{mn}^\dag$ are two fermionic coordinates.
  The theory has $SU(N)$ symmetry as well as a flavor $SO(3)$ symmetry.
  We do not assume
  supersymmetry, and so the choice of variables as well as the form of the
  Lagrangian is not dictated by symmetry.  Instead, we took a bottom
  up approach in the art of phenomenological model
  building.  The relative coefficient
  between  the mass term and the commutator square term
  has been choosen to be a factor of 4.
  This can be achieved by rescaling the $X$'s.
  The Lagrangian is characterized by a single mass scale
  $M_P$ called the Planck mass, which appears in front of the
  commutator square term with a unit coefficient. The Yukawa coupling
  and  bosonic kinetic term have dimensionful coefficients which we specifiy
  in terms of $M_P$ through the coefficient $a_2$ and $a_0$ where
  $M_0 = a_0^2 M_P$.
  In addition, we have included a topological action term specified by $a_3$.
Here $r_X$ is the rank of the matrix $\G:= [X^a, X^b]^2$. The $r_X$ term
does not affect the equation of motion, but measures the energy of space
due to noncommutativity.  
We have shown in \cite{Chu:2024qil} that the
reproduction of the energy and entropy of
the Schwarzschild black hole requires to identify
the Planck mass and the  Planck length  in terms of
  the Newtonian constant $G$ as
\be
M_P = \frac{1}{\g}\sqrt{\frac{2}{\pi G}}, \quad 
l_P =  \sqrt{\frac{2G}{\pi}} ,
\ee
where $\g := a_2-1+ 2a_3 $ and the coefficients $a_2, a_3$
are constrainted such that $\g >0$. In this paper, we show that
these relations are unmodified (as required) and the rotating fuzzy sphere
solution
of the model matches well the shape, size, energy, entropy and the
angular momentum of the Kerr black hole.

Let us consider dynamical solution to the theory.
The classical equation of motion for a bosonic matrix configuration 
is given by
   \be \label{eom}
  -\ddot{X}^a  + \frac{4a_0^2 M_P^2}{ N^2}
   \left( [X^b,[X^a,X^b]] + 2 X^a \right) =0.
   \ee
   The Kerr black hole is a stationary solution of general relativity with
   axial symmetry.
   Suppose the rotation is around the $z$-axis, then it is natural to consider
   a new basis
of coordinates $(X^+, X^-, X^3)$ where
\be
X^\pm := \frac{1}{\sqrt{2}}(X^1 \pm i X^2).
\ee
In the new basis, the equations of motion reads
\bea
 - \ddot{X}^+ + \frac{4a_0^2 M_P^2}{N^2}
 \left([X^+,[X^+, X^-]] +  [X^3,[X^+, X^3]]
+ 2 X^+  \right) =0, \label{eom1}\\
- \ddot{X}^3 + \frac{4a_0^2 M_P^2}{N^2}\left([X^-,[X^3, X^+]] +  [X^+,[X^3, X^-]]
+ 2 X^3 \right)=0.
\label{eom2}
\eea
The equation of motion for $X^-$ is not independent and can be
obtained from the one \eq{eom1} of  $X^+$. For rotation, let us
consider  the  ansatz
\be
X^\pm(t) = e^{\pm i \o t} X^\pm (0), \quad  \mbox{$X^3$ independent of $t$}
\ee
which is appropriate for axial symmetric rotational motion.
The equation of motion becomes
\bea \label{eom2a}
   [X^+,[X^+, X^- ] ] +  [X^3,[X^+, X^3] ] + 2 c^2 X^+ =0,\;\; \\
    \label{eom2b}
   [X^-, [X^3,X^+ ] ] + [X^+,[X^3,X^- ] ] +2X^3 =0,\;\;
\eea
where $c^2 := 1+N^2 \o^2/(8 a_0^2 M_P^2)$
and the equations are written for time $t$.
The equations \eq{eom2a} and \eq{eom2b}
can be solved by utilizing the spin $j=(N-1)/2$ representation of  $SU(2)$
in the Cartan-Weyl basis. They satisfy the commutation relations
\be \label{T1}
[T^+, T^-] = T^3, \quad [T^3, T^\pm] = \pm T^\pm
\ee
and the Casimir relation
\be \label{T2}
T^+ T^- + T^- T^+ + T_3^2 = \frac{N^2-1}{4} \id.
\ee
It is easy to see that
\be \label{X-fuzzy}
X^\pm (t) = e^{\pm i \o t} T^\pm, \quad
X^3(t) = c_3 T^3,\quad
c_3 = \sqrt{1+\frac{N^2\o^2}{4a_0^2 M_P^2}}
\ee
solve the equation of motion. As $c_3 \geq 1$, we can introduce the definition
\be \label{c3}
\cos \b = 1/c_3,
\ee
which is a measures of the angular velocity.
Let us next introduce the dimensional coordinates. 
To motivate the non-commutative
version of \eq{cc}, note that the matrices 
\be \label{dc}
\hat{T}^a := \frac{2}{\sqrt{N^2-1}} T^a
\ee
 plays  the role of the directional cosines 
 $(\sin \th \cos \phi, \sin \th \sin \phi, \cos \th)$ in the non-commutative case. 
As a result, we adopt 
\be \label{YY0}
Y^\pm = 2l_P X^{\pm},\quad Y^3 = 2 l_P X^3/c_3^2
\ee
as the dimensionful Cartesian coodinates in the oblate spherical coordinate form.
Here the normalization factor has been choosen such that, when written in terms of 
the "directional cosines" \eq{dc}, we have  
\be \label{YY}
Y^\pm = l_P \sqrt{N^2-1} \;  e^{\pm i \o t} \hat{T}^\pm, \quad
Y^3 = l_P \sqrt{N^2-1} \cos \b \; \hat{T}^3.
\ee 
This agrees with the form of \eq{cc} if the parameters $R$ and $a$ are extracted from our solution as
\be \label{Ra}
 \sqrt{a^2 +R^2} = l_P \sqrt{N^2-1}, \quad R = l_P \sqrt{N^2 -1} \cos \b
 \ee
 and note the fact that our solution corresponds to the rotation $\phi = \o t$.
In  terms of $Y$'s, our solution  satisfies the  commutation relations
\be \label{fe2}
   [Y^+, Y^-] = \frac{2R/\cos \b }{ \sqrt{N^2-1}} Y^3,
    \quad
        [Y^3, Y^\pm] = \pm  \frac{2R \cos \b}{ \sqrt{N^2-1}}Y^\pm
   \ee
and the constraint
\be \label{fe1}
\frac{Y_1^2 + Y_2^2}{R^2 +a^2} + \frac{Y_3^2}{R^2} =1.
\ee 
Note that the relations \eq{fe1}, \eq{fe2}
   are time independent and they define an oblate fuzzy
   ellipsoid for $\o \neq 0$.
It is interesting that the shape \eq{fe1} of our solution 
agrees precisely with \eq{kerr-shape} of the Kerr black hole.
This provides a first check of our claim that the rotating fuzzy ellipsoid solution
provides a description of the quantum Kerr black hole.
Note also that   \eq{Ra} takes on the simple form
\be \label{Ra1}
R =  N l_P\cos \b, \quad a =  N l_P \sin \b
\ee
in   the leading order of large $N$ limit, which we will assume from now on. 

Over the geometry of the fuzzy ellipsoid, the Hamiltonian of the system
reads
\be
H = H_B + H_F,
\ee
where $H_B$ is the bosonic 
part of the Hamiltonian and $H_F$ is the part of the Hamiltonian. It is 
\be \label{HB}
H_B = (2a_3 -1)\frac{N M_P}{2} + \frac{N^3 \o^2}{12 a_0^2 M_P},
\ee
where we have used $\tr T_3^2 = \frac{1}{3} \tr T^2 =  \frac{N^3}{12}$.
The fermionic Hamiltonian
$H_F =  a_2 \frac{M_P}{N^2}\psi^\dag \s^a X^a \psi $
is more interesting. Due to the anisotropy introduced by the coordinates
\eq{X-fuzzy}, we can write
\be
H_F = H_F^0 + h_F,
\ee
as  an isotropic part 
\be \label{HT0}
H_F^0 = a_2 \frac{M_P}{N^2} \psi^\dag \s^a T^a(t) \psi
\ee
plus a term $h_F$
\be \label{pert}
h_F =  a_2 (c_3 -1) \frac{M_P}{N^2} \psi^\dag \s^3 T^3 \psi :=
\psi^\dag \tilde{h}_F \psi,
\ee
where in \eq{HT0},
\be \label{Tt}
T^\pm(t) := e^{\pm i \o t} T^\pm, \quad T^3(t) := T^3
\ee
denote the generators of $SU(2)$ under a rotation of angle $\o t$
around the $z$-axis; and  we have introduced the matrix
\be
\tilde{h}_F:=  a_2 (c_3 -1) \frac{M_P}{N^2} \s^3 T^3
\ee
in \eq{pert}. 
The term $h_F$ is treated as a perturbation and is of the order of
$a^2/R^2$ for slow rotation. The unperturbed
Hamiltonian $H_F^0$ can be diagonalized as in \cite{Chu:2024qil} and
give rises to a Fermi sea.
In fact, the matrix  $K := \s^a T^a(t)$ admits the  eigendecomposition 
\be \label{K-eigen}
K_{(m\a) (n\b)} = \frac{N}{2} \sum_{p=1}^{N}
\left(\cU^{p}_{m\a} \cU^{p\dag}_{n\b} - \cV^{p}_{m\a} \cV^{ p \dag}_{n\b}
  \right),
  \ee
where  $\cU^{p}_{n\b}, \cV^{p}_{n\b} \; (p =1, \cdots, N)$ are
eigenvectors of $K$, which now become time-dependent. They satisfy the
orthonormality condition and the completeness relation
  \bea
&& \cU^{p\dag}_{m\a} \cU^q_{m\a} = \cV^{p\dag}_{m\a} \cV^q_{m\a} = \d^{pq}, \qquad 
\cU^{p\dag}_{m\a} \cV^q_{m\a} =0, \\
&& \cU^{p}_{m\a} \cU^{p\dag}_{n\b} +\cV^{p}_{m\a} \cV^{p\dag}_{n\b} =
\d_{mn} \d_{\a\b},
\eea
which are however time independent and identical to before.
Introducing  the fermionic oscillators
  \be \label{xi-chi}
  \xi^{p}_k := \cU^{p \dag}_{n\b} \psi_{nk\b}, \quad
   \chi^{p\dag}_k := \cV^{p \dag}_{n\b} \psi_{nk\b}, 
   \ee
and after subtracting away a constant due to normal ordering,
$H_F^0$ can be written in  the diagonal form
\be \label{HF-diag}
   H_F^0 =  a_2 \frac{M_P}{2N}
     \left(\sum_{p,k=1}^N 
     \xi_k^{p\dag} \xi_k^{p}+ \chi_k^{p\dag} \chi_k^{p}\right)
     \ee
and has the  eigenstates
\be \label{es}
\left|\Psi^{p_1 \cdots p_r q_1 \cdots q_s}_{k_1 \cdots k_r l_1 \cdots l_s}\right\rangle
:=\xi^{p_1\dag}_{k_1}\cdots \xi^{p_r\dag}_{k_r} \chi^{q_1\dag}_{l_1} \cdots
\chi^{q_s\dag}_{l_s} \ket{0},
\ee
where $\ket{0}$ is the fermionic Fock vacuum. 
Denoting $r, s$ the occupation number for the $\xi$-type and $\chi$-type
fermionic oscillators respectively, it reads
\be
H_F^0 =  a_2\frac{M_P}{2N} (n+N^2), \quad n := r+s - N^2,
\ee
where $n = -N^2, \cdots, 0, \cdots, N^2$
specifies the energy level within the Fermi sea.
The lowest level $n = -N^2$ corresponds
to  an empty Fermi sea,
while the highest level $n=N^2$ corresponds to
a completely filled Fermi sea.

We can divided the space of eigenstates of $H_F^0$
according to their corresponding
level $n$.  In general, the set of
eigenstates at level $n$ is given by
\be
\mathbb{V}_n:= \{\ket{i}, i= 1, \cdots, \Omega_n\},
\quad \Omega_n = \binom{2N^2}{N^2 + n}
\ee
where $\ket{i}$
are those states \eq{es} with $r+s = N^2 +n$ and 
$\Omega_n$ is the 
total number of states at the level $n$.
As in \cite{Chu:2024qil}, one see that
the Bekenstein entropy  bound \cite{Bekenstein:1980jp} is satisfied
for the fuzzy 
configurations with $n \geq 0$
and the bound is saturated at $n=0$.  This suggests that the $n=0$
configuration,
being most compactly packed, should be identified with a black hole.
Let us therefore consider the configuration
of a half-filled  ($n=0$) Fermi sea. We have 
\be
\Omega_0 = 2^{2N^2}
\ee
in the leading order of large $N$.
Since classically we do not make observation of the quantum states,
this give rises to a coarse grained entropy
$S= \log_2 \Omega_0$: 
\be \label{S-QM}
S = 2N^2 = \frac{\pi (R^2+a^2)}{G}.
\ee
This is precisely the Bekenstein-Hawking entropy of a Kerr black hole
if we identify $R$ with $r_+$ of the outer horizon radius of the Kerr
black hole. To make this identification, we need to show that $R$
satisfies the relation \eq{M-GR} for a given mass $M$ of the
Kerr black hole. For this, we need to include the energy of the system
of Fermi oscillators in the fuzzy background,
which we will do next. 

For the half-filled
Fermi sea, we have
\be \label{E-HFS}
H_F^0 =  a_2\frac{M_PN}{2}
\ee
and we are left with 
the contribution to the energy due to the  perturbation $h_F$.
On this, we
may use the perturbation theory
for degenerate energy level to compute the correction to the $n=0$ energy
level.
This requires the
knowledge  of the matrix elements
$\langle i | h_F | j \rangle $ and
it's eigenvalues, where
$\ket{i}$ 
are the unperturbed states in $\mathbb{V}_0$.
However, physically the set of microstates is not observed. It is
more meaningful to consider the average corrections over the set of microstates.
Since the system is in isolation, the set of microstates form a
microcanonical ensemble. It was shown in the appendix of
\cite{Chu:2024qil} that  the ensemble average  of the corrections
\be \label{en-def}
\langle h_F \ra_{\rm En} := \frac{1}{\Omega_0} \sum_{\ket{i} \in\mathbb{V}_0} 
\bra{i} h_F \ket{i}
\ee
is given by
\be \label{en-ave}
\la h_F \ra_{\rm En}  = \tr(K \tilde{h}_F).
\ee
This holds in general for any form of $\tilde{h}_F$.
For the present case with \eq{pert},
only the $T^3$ component of $K$ contributes and so we get
\be \label{hF-en}
\la h_F\ra_{\rm En}   = a_2 \frac{c_3-1}{6} N M_P.
\ee
We remark that in the current case of a fuzzy ellipsoid background,
the eigenvalues of the fermionic Hamiltonian and  hence  the ensemble average
\eq{en-def}
can  be determined exactly; and  \eq{hF-en}
can be obtained consequently from a slow rotation expansion.
We will demonstrate this
in the appendix of this paper.
Adding
\eq{E-HFS}, \eq{hF-en} to \eq{HB}
and expressing it in terms of macroscopic quantities $R$
and $a$, we have
\be \label{E1}
E = \frac{R^2+a^2}{2 GR} f(\b), \quad
f (\b) := \cos \b \left(1+\frac{2d}{3} \tan^2 \b \right),
\ee
where $d := (1+ {a_2}/{4})/\g$.
It is $f=1$ for $\b=0$.
For $\b \ll 1$, $f$ admits an expansion in terms of $\tan^2 \b$.
This corresponds to an expansion of small angular momentum or small
curvature since $\tan^2 \b = a^2/R^2$.
In order not to reproduce the Kerr mass, considered as a
small  ${a^2}/{R^2}$ expansion of the Schwarzschild
mass, the leading term $\tan^2 \b$ of $f(\b)$ in \eq{E1}
should vanish. This happens
when $d=3/4$, or equivalently
\be \label{a2a3}
a_2+ 3 a_3  = \frac{7}{2}.
\ee
This implies that $f$ has the small $\b$ expansion
\be \label{f-exp}
f = 1 + \frac{1}{8} \tan^4 \b + \cdots. 
\ee
and we obtain  the result \eq{M-GR} of general relativity (GR)
\be \label{E2}
M = \frac{R^2+a^2}{2 GR}
\ee
in the leading order of small $a^2/R^2$.
It is remarkable that the Yukawa term in the Lagrangian \eq{L}
not only give rises to the needed microstates to account for the
entropy of quantum black hole, but also makes contribution
to the energy of the fuzzy system, without which
the mass of the GR Kerr black hole would not be obtained.
We note  that in the case of the Schwarzschild black hole
\cite{Chu:2024qil}, the Hamiltonian over the fuzzy geometry can be
treated exactly and our large $N$ quantum mechanics reproduces
precisely the GR result for the Schwarschild mass. On the contrary for
the present case of rotating black hole, a perturbative treatment is
needed to handle the Hamiltonian over the fuzzy ellipsoid geometry and
the result \eq{E2} was obtained at the order of small $a^2/R^2$. To
check it at the next order, the second order pertutative result and
the $\sin^4 \b$ contribution of \eq{f-exp} should add up to zero at
the order of $a^4/R^4$.  In general, to reproduce \eq{E2} for
arbitrary angular momentum $a/R$, some nontrivial cancellation among
different orders of perturbation theory should happen.  This suggests
the presence of integrability \cite{Berenstein:2002jq} and is a
nontrivial test of our proposal.

The Kerr black hole also carries an angular momentum \eq{J-GR}. To facilitate
the comparsion with the quantum mechanics prediction, let us rewrite it in
terms of the matrix model quantities. Using \eq{Ra1} and \eq{c3}, we obtain
\be \label{JGR}
J_{GR} = \frac{\o N^3}{2 \pi a_0 M_P}.
\ee
Let's check it
against our rotating solution. In the quantum mechanics, the rotational
symmetry give rises to the angular momentum generators
\be
L^a = \e_{abc} \tr \; X^b P^c, \quad S^a = \tr \; \psi^\dag \frac{\s^a}{2} \psi.
\ee
For the rotating fuzzy sphere solution,
it is easy to obtain
\be \label{L3}
L^3 = \frac{\o N^3}{6 a_0^2 M_P}, \quad L^{1,2} =0.
\ee
As for the spin, since the half-filled Fermi sea is not observed in GR,
we consider the ensemble average $\la S^a \ra_{\rm En}$. It is then easy to use
\eq{en-ave} for $\tilde{h}_F = \s^a/2$ to obtain
\be
\la S^a \ra_{\rm En} =0.
\ee
Therefore the angular momentum of the fuzzy sphere is entirely given by the
orbital $L_3$. It is interesting to note that \eq{JGR} and \eq{L3} takes has the
same dependence in $N$ and a precise agreement is possible if
\be
a_0 = \frac{\pi}{3} .
\ee
All in all, the proposed quantum mechanics with
$a_0 = \pi/3$ and \eq{a2a3} contains solutions that match
well with the GR description of the Schwarzschild and the Kerr black holes.
In addition the black hole entropy is reproduced by a microstate counting.
It is important to find out if the Newtonian potential between Schwarzschild
black holes can be reproduced with a further constraint on the coefficients
of the model. 

Finally, we finish the paper with some discussions. 
We remark that
in our quantum mechanics, we have, for finite $N$, quantum spaces and 
quantum gravity corrections are given by an $1/N$ expansion.
It is interesting to understand how the large $N$ expansion
give rises to the low energy effective expansion of gravity and beyond.
It is also interesting to explore if there is some noncommutative geometric
way to express these quantum corrections.
In string theory, the entropy of certain class of extremal black hole has 
been reproduced beautifully by a counting of BPS states
in a weakly coupled gauge theory 
\cite{Strominger:1996sh}. 
There the gauge system becomes a
black hole in the strong coupling regime and the
counting  is protected by supersymmetry.
Our counting is different. 
We do not rely on the use of supersymmetry or duality.
Instead, we employ noncommutative geometry to model quantum gravity
and  we counted   the fermionic microstates 
that makes up the  black hole.  
Let us also comment on the 
black hole thermodynamics. 
Although our fuzzy solutions are classically stable, 
quantum mechanically there can be tunneling between them. We
suspect that 
this gives the Hawking radiation
and give rises to the black hole thermodynamics.
There are also other interesting problems. 
It is interesting to studying the  form factor 
\cite{Maldacena:2001kr,Papadodimas:2015xma} 
for our quantum mechanics
and study how chaotic behaviour of black hole arises in the large $N$ limit. 
It is interesting to exact spin dependent force between black holes from
the quantum mechanics.
It is also interesting to think about how the ergosphere of Kerr
black hole appears in our description. A probe analysis may help to
reveal it.
We leave  these interesting problems  for future consideration.

\section*{Acknowledgments}
The support of this work by NCTS and
the grants 110-2112-M-007-015-MY3 and 113-2112-M-007-039-MY3
of the National 
Science and Technology Council of Taiwan is gratefully acknowledged. 

\appendix

\section{Ensemble average from the exact eigenvalues}
In the maintext of the paper, we have obtained the ensemble average
\eq{hF-en} for the Hamiltonian $h_F$ by treating it as a perturbation.
In this appendix, we show that the eigenvalues of the total
Hamiltonian $H_F = H_F^0+h_F$ can be determined exactly and the result
\eq{hF-en} can be obtained from an average of $H_F$
over the ensemble of microstates.

Consider in general
adjoint fermions $\psi_{mn \;\a}$ of $SU(N)$ and a Hamiltonian
\be
H := \tr \;  \psi^\dag \Ht\psi,
\ee
where $(\Ht)_{(m\a) (n\b)} $  is an arbitrary
matrix. 
Suppose $\Ht$ admits the eigenvectors $\cU^{p_+}, \cV^{p_-}$ that corresponds
to positive and negative eigenvalues $\l_{p_+}, \l_{p_-}$ respectively. We
have the eigen-decomposition of $\Ht$
\be \label{eigen-X}
(\Ht)_{(m\a) (n\b)} =
\sum_{p_+} \l_{p_+} \cU^{p_+}_{m\a} \cU^{p_+\dag}_{n\b}
+ \sum_{p_-} \l_{p_-}\cV^{p_-}_{m\a} \cV^{ p_- \dag}_{n\b} .
\ee
The eigenvectors satisfy the orthonormality condition
  \be
  \cU^{p_+\dag}_{m\a} \cU^{q_+}_{m\a} = \d^{p_+ q_+}, \quad
  \cV^{p_-\dag}_{m\a} \cV^{q_-}_{m\a} = \d^{p_-q_-}, \qquad 
\cU^{p_+\dag}_{m\a} \cV^{q_-}_{m\a} =0
\ee
and the completeness relation
\be
\sum_{p_+}\cU^{p_+}_{m\a} \cU^{p_+\dag}_{n\b} +
\sum_{p_-}\cV^{p_-}_{m\a} \cV^{p_-\dag}_{n\b} =
\d_{mn} \d_{\a\b}.
\ee
Introduce the fermionic oscillators defined by
  \be \label{xi-chi-a}
  \xi_k^{p_+} := \cU^{p_+ \dag}_{n\b} \psi_{n k \b}, \quad
   \chi_k^{p_-\dag} := \cV^{p_- \dag}_{n\b} \psi_{n k\b},
   \ee
$H$ takes the diagonal form
\be
H =
\sum_{p_+,k} \l_{p_+} \xi_k^{p_+\dag} \xi_k^{p_+} -
\sum_{p_-,k} \l_{p_-} \chi_k^{p_-\dag} \chi_k^{p_-}.
\ee
Here we have choosen a normal ordering prescription such that
$\mathop{:} \chi \chi^\dag \mathop{:}  = - \chi^\dag \chi$ etc.
$H$ has the eigenstates
\be \label{es-n}
\left|\Psi^{p_1 \cdots p_r q_1 \cdots q_s}_{k_1 \cdots k_r l_1 \cdots l_s}\right\rangle
:=\xi^{p_1\dag}_{k_1}\cdots \xi^{p_r\dag}_{k_r} \chi^{q_1\dag}_{l_1} \cdots
\chi^{q_s\dag}_{l_s} \ket{0},
\ee
where $\ket{0}$ is the Fock vacuum.
Let us consider the set of states \eq{es-n} 
that corresponds to a half-filled Fermi sea (HFS):
\be
\mathbb{V}_{\rm HFS}:= \{\ket{i}, i= 1, \cdots, \Omega_0\},
\ee
where $\ket{i}$
are those states \eq{es-n} with $r+s = N^2$
and $ \Omega_0= \binom{2 N^2}{ N^2}$ is the number of states that satisfy
this condition. We are interested in the ensemble average defined by
\be
\la f \ra_{\rm En}: = \frac{1}{\Omega_0} \sum_{\ket{i} \in\mathbb{V}_0} 
\bra{i} f \ket{i}.
\ee
As a result, we have
\be 
\la H \ra_{\rm En}  = 
\sum_{p_+} \l_{p_+} n(p_+)  -
\sum_{p_-} \l_{p_-} n(p_-), 
\ee
where $n(p_\pm)$ denotes the occupation number for the eigenvalue level
$p_{\pm}$:
\be
 n(p_+) := \sum_{k} \la \xi_k^{p_+\dag} \xi_k^{p_+}\ra_{\rm En} ,\quad
 n(p_-) := \sum_{k} \la \chi_k^{p_-\dag} \chi_k^{p_-}\ra_{\rm En}.
\ee
For an isolated system, the occurance probability is the same for all energy
levels, therefore
\be
n(p_+) = n(p_-) = \frac{N}{2}.
\ee
This is obtained by recalling that the total number of $p_+$ and $p_-$ is $2N$
in the large $N$ limit and that we have the constraint
\be
\sum_{p_+} n(p_+) + \sum_{p_-} n(p_-) = N^2
  \ee
  for a system of microstates that filled half of the Fermi sea.
  As a result, we obtain
\be\label{H-en}
\la H \ra_{\rm En}  = \frac{N}{2}\;
(\sum_{p_+} \l_{p_+} - \sum_{p_-} \l_{p_-}).
\ee
This result holds generallly independent of the form $\Ht$
of the Hamiltonian. For our QM,
$\Ht =\frac{a_2 M_P}{N^2} \s^a X^a$.
For the single fuzzy sphere background,
$\l_{p_+} = - \l_{p_-} = \frac{a_2 M_P}{N^2} \cdot \frac{N}{2}$ and
so $\la H \ra_{\rm En} = a_2 M_P N/2$.
For the displayed fuzzy spheres background, the same result is obtained
 since the relative displacement of the fuzzy
 spheres induces a magnetic dipole term whose ensemble average is zero
 \cite{Chu:2024qil}.

 Let us now consider the rotating fuzzy sphere system. In this case,
 consider  $K = \s^a T^a(t) +(c_3-1) \s_3 T^3$ where
 $T^a$ are the spin $j=(N-1)/2$
 representation of $SU(2)$ as specified by \eq{T1},
 \eq{T2} and $T^a(t)$ as defined in \eq{Tt}
 denotes the generators of $SU(2)$ under a rotation. It is easy to see that
 $K$ is diagonalized by the angular momentum states $\ket{j, m}$,
 $m = -(N-1)/2 \cdots, (N-1)/2$. We obtain the eigenvalues:
 \be \label{eigen-v}
 \l_+ = c_3 \frac{N-1}{2} \quad\mbox{and}\quad
 -\frac{c_3}{2} + \frac{1}{2}\sqrt{N^2+ (c_3^2-1)(2m+1)^2},
 \quad m = -\frac{N-1}{2} , \cdots, \frac{N-1}{2},
 \ee
 \be
\l_- = -\frac{c_3}{2} - \frac{1}{2}\sqrt{N^2+ (c_3^2-1)(2m-1)^2}, \quad
 \quad m =  -\frac{N-1}{2}+1 , \cdots, \frac{N-1}{2}.
 \ee
 There are in total $N+1$ positive eigenvalues ($\l_+$)
 and $N-1$ negative ones ($\l_-$).
 Note that
 \be
\sum \l_+ + \sum \l_- =0, 
\ee
reflecting the fact that $K$ is traceless.
Note also that
 in the undeformed case $c_3=1$, the eigenvalues reduce to their correct
 limit: $\l_+ = (N-1)/2$, $\l_- = -(N+1)/2$.

With the set \eq{eigen-v} of eigenvalues, we can now compute the ensemble
value $\la H \ra_{\rm En} $ using \eq{H-en}. Expanding \eq{eigen-v}
in the limit of small $(c_3-1)$, we obtain in the leading
limit of large $N$ and small $(c_3-1)$ ,
\be
\sum \l_+ - \sum \l_- = N^2 + (c_3-1) \frac{N^2}{3}.
\ee
Reinserting back the
proportional coefficient $\frac{a_2 M_P}{N^2}$, we final obtain
\be
\la H \ra_{\rm En} = \frac{a_2 M_P N}{2} + a_2 M_P N \frac{c_3-1}{6}.
\ee
This agrees precisely with \eq{E-HFS} and \eq{hF-en}, in which the latter
was obtained from \eq{en-ave}
without computing the exact eigenvalues.

\bibliography{references}

\end{document}